\documentclass[twocolumn,british,sort&compress]{article}

\usepackage[T1]{fontenc}
\usepackage[latin9]{inputenc}
\usepackage{geometry}
\usepackage{color}
\usepackage{babel}
\usepackage{verbatim}
\usepackage{amsbsy}
\usepackage{amstext}
\usepackage{amssymb}
\usepackage{graphicx}
\usepackage{wasysym}
\usepackage{esint}
\usepackage[super,sort&compress,comma]{natbib}
\usepackage[unicode=true,pdfusetitle,
 bookmarks=true,bookmarksnumbered=false,bookmarksopen=false,
 breaklinks=true,pdfborder={0 0 0},pdfborderstyle={},backref=false,colorlinks=true]
 {hyperref}
\hypersetup{
 citecolor=blue,urlcolor=blue,linkcolor=blue}

\makeatletter
\newcommand{\lyxaddress}[1]{
\par {\raggedright #1
\vspace{1.4em}
\noindent\par}
}

\usepackage{dblfloatfix}    
\usepackage{doi}
\date{ }

\newcommand*{\citen}[1]{%
	\begingroup
	\romannumeral-`\x 
	\setcitestyle{numbers}%
	\cite{#1}%
	\endgroup   
}

\makeatother

\begin{document}

\title{Phase-matched extreme-ultraviolet frequency-comb generation}

\author{Gil Porat\textsuperscript{1{*}}, \and Christoph M. Heyl\textsuperscript{1,2{*}},
\and Stephen B. Schoun\textsuperscript{1{*}}, \and Craig Benko\textsuperscript{1},
\and Nadine Dörre\textsuperscript{3}, \and Kristan L. Corwin\textsuperscript{1,4},
\and Jun Ye\textsuperscript{1}}
\maketitle

\lyxaddress{\textsuperscript{1}JILA, NIST and the University of Colorado, 440
UCB, Boulder, Colorado 80309-0440, USA\\
\textsuperscript{2}Department of Physics, Lund University, P.O. Box
118, SE-221 00 Lund, Sweden\\
\textsuperscript{3}University of Vienna, Faculty of Physics, VCQ,
Boltzmanngasse 5, A-1090 Vienna, Austria\\
\textsuperscript{4}Department of Physics, Kansas State University,
116 Cardwell Hall, Manhattan, Kansas 66506, USA\\
\textsuperscript{{*}}These authors contributed equally to this work.}

\textbf{Laser-driven high-order harmonic generation~\citep{McPherson1987,FerrayJPB1988}
(HHG) provides tabletop sources of broadband extreme-ultraviolet (XUV)
light with excellent spatial~\citep{Bartels2002} and temporal~\citep{Benko2014}
coherence. These sources are typically operated at low repetition
rates, $\boldsymbol{f_{\mathrm{rep}}\lesssim100\;\mathrm{kHz}}$,
where phase-matched frequency conversion into the XUV is readily achieved~\citep{Constant1999,Paul2006}.
However, there are many applications that demand the improved counting
statistics or frequency-comb precision afforded by operation at high
repetition rates, $\boldsymbol{f_{\mathrm{rep}}>10\;\mathrm{MHz}}$.
Unfortunately, at such high $\boldsymbol{f_{\mathrm{rep}}}$, phase
matching is prevented by the accumulated steady-state plasma in the
generation volume~\citep{Yost2011,Allison2011,Carlson2011,Lee2011,Mills2012},
setting stringent limitations on the XUV average power. Here, we use
gas mixtures at high temperatures as the generation medium to increase
the translational velocity of the gas, thereby reducing the steady-state
plasma in the laser focus. This allows phase-matched XUV emission
inside a femtosecond enhancement cavity at a repetition rate of 77~MHz,
enabling a record generated power of ${\boldsymbol{\sim}}$2~mW in
a single harmonic order. This power scaling opens up many demanding
applications, including XUV frequency-comb spectroscopy~\citep{Cingoz2012,Kandula2010}
of few-electron atoms and ions for precision tests of fundamental
physical laws and constants~\citep{Drake2008,Eyler2008,Herrmann2009,Karshenboim2005,Palffy2010,Ubachs2014,Vogel2013}. }

The highly-nonlinear HHG process requires peak laser intensities around
$10^{14}~\textrm{W/cm}^{2}$, which necessitates large laser pulse
energies $\gtrsim10\;\mu\mathrm{J}$, and short pulse durations $\lesssim100$~fs,
as typically reached with low repetition rate, chirped-pulse amplified~\citep{Backus1998}
laser systems. However, high repetition rates are desirable for applications
such as photoelectron spectroscopy~\citep{Chiang2012,Frietsch2013,WallauerAPL2016}
and microscopy \citep{StockmanNatPho2007} as well as electron-ion
coincidence spectroscopy~\citep{Dorner2000,Sabbar2014}, which are
limited by counting detection or space-charge effects to few XUV ionization
events per shot. Most notably, precision frequency-comb spectroscopy~\citep{Cingoz2012,Kandula2010}
requires $f_{\mathrm{rep}}\gg10\,\mathrm{MHz}$ in order to stabilize
the comb. Recent efforts allowed HHG to be directly driven at $f_{\mathrm{rep}}\gtrsim1\,\mathrm{MHz}$,
using either the direct output of a high-power oscillator \citep{Chiang2012,EmauryOpt2015}
or the coherent combination of several fibre amplifiers \citep{Hadrich2015,Rothhardt2014a}.
Achieving the necessary intensity for HHG with $f_{\mathrm{rep}}\gg10\,\mathrm{MHz}$
requires lasers with average power in the kW range. Apart from one
demonstration at 20~MHz, where the measured XUV power was extremely
low~\citep{VernalekenOL2011}, higher repetition rates up to 250~MHz~\citep{Carstens2016}
have been facilitated only by using passive enhancement cavities,
which store $\text{\ensuremath{\sim}10}$ kW of laser power, where
a gas jet is introduced at an intracavity focus~\citep{Gohle2005,Jones2005,Cingoz2012,Yost2011,Lee2011,Mills2012,Pupeza2013}.

\begin{figure*}[t]
\begin{centering}
\includegraphics[width=1\textwidth]{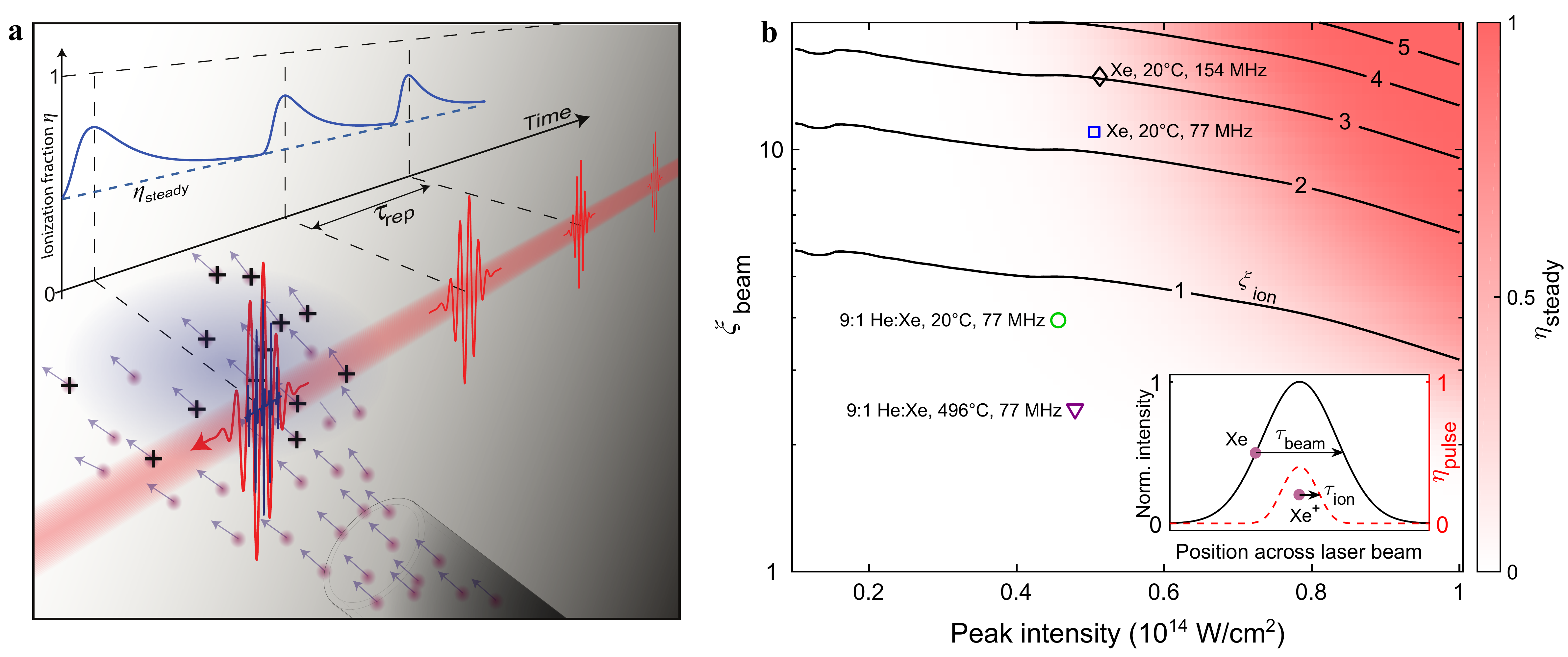}
\par\end{centering}
\caption{(a) Overview of plasma dynamics in high-repetition-rate high-harmonic
generation. A train of femtosecond laser pulses crosses and ionizes
a xenon gas jet. The interval between consecutive pulses, $\tau_{\mathrm{rep}}=1/f_{\mathrm{rep}}$,
is smaller than the plasma transit time through the ionization volume,
resulting in a high steady-state ionization fraction, $\eta_{\mathrm{steady}}$.
(b) Spatially-averaged steady-state ionization fraction, $\eta_{\mathrm{steady}}$
(colour scale), as a function of peak laser intensity and the number
of laser pulses that cross the gas jet during the gas transit time
through the laser beam, $\xi_{\mathrm{beam}}=\tau_{\mathrm{beam}}/\tau_{\mathrm{rep}}$.
The contour lines show the intensity-dependent number of laser pulses
$\xi_{\mathrm{ion}}=\tau_{\mathrm{ion}}/\tau_{\mathrm{rep}}$ that
cross the gas jet during the time it takes an ion to clear the ion-generation
volume, as illustrated in the inset. The black diamond ($\eta_{\mathrm{steady}}=17\%$),
blue square ($\eta_{\mathrm{steady}}=11\%$), green circle ($\eta_{\mathrm{steady}}=1.1\%$),
and purple triangle ($\eta_{\mathrm{steady}}=0.2\%$), indicate experimental
conditions of optimal 11\protect\textsuperscript{th} harmonic yield
for various gas and laser parameters.\label{fig:cartoon}\label{fig:eta_steady}}
\end{figure*}

In a macroscopic extended medium, efficient HHG requires matching
the phase velocities of the generating laser and the generated fields.
This can be achieved by balancing neutral and plasma dispersion, the
geometric phase shift due to focusing (the Gouy phase), and the HHG
intrinsic dipole phase \citep{Constant1999,Popmintchev2010}. Achieving
this balance becomes increasingly challenging as the repetition rate
increases above ${\sim}$10~MHz. The reason for this difficulty is
that the plasma generated by one pulse does not have time to clear
the focal volume before the next laser pulse arrives and generates
even more plasma. Consequently, a high-density steady-state plasma
is formed~\citep{Allison2011,Carlson2011}, which is highly dispersive
(see Fig.~\ref{fig:cartoon}a), making phase matching unattainable.
While phase-matched HHG has been demonstrated at a repetition rate
of 10.7 MHz \citep{Hadrich2015}, it has not been previously achieved
at higher repetition rates, to the best of our knowledge.

\begin{figure*}[t]
\begin{centering}
\includegraphics[width=1\textwidth]{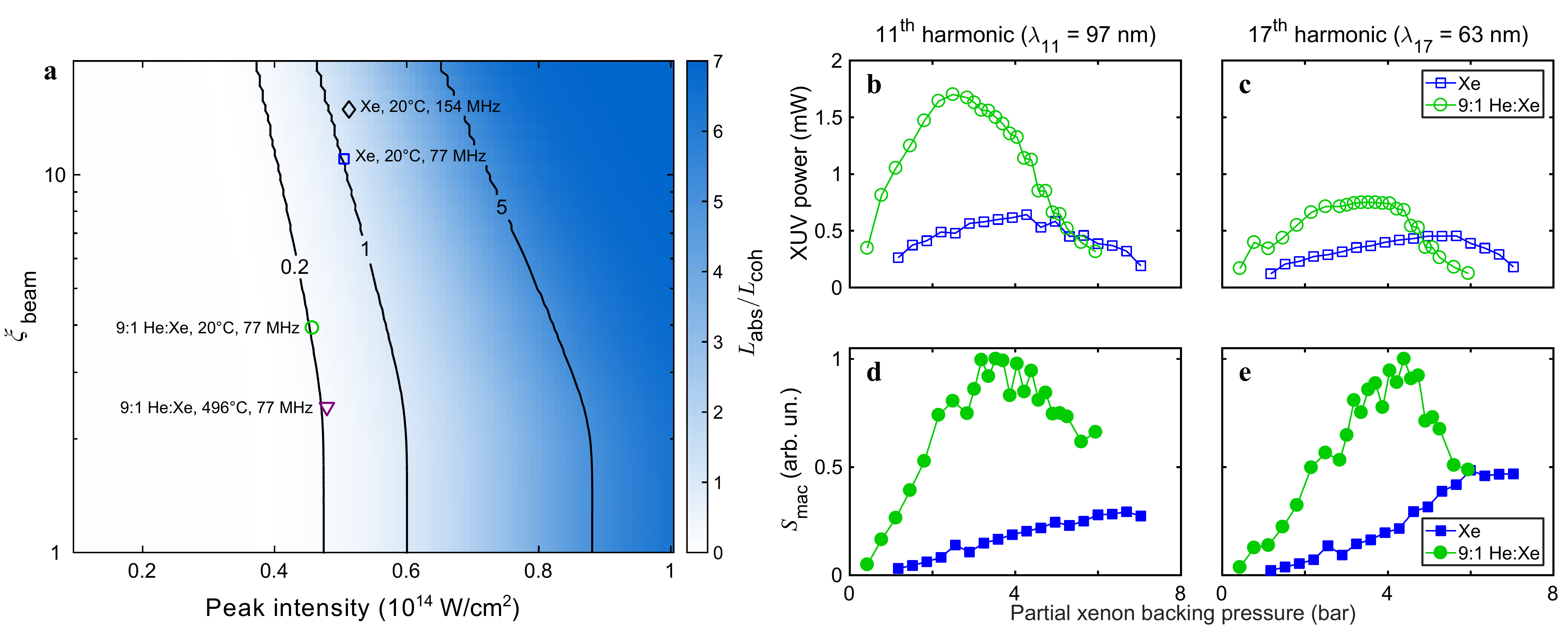}
\par\end{centering}
\caption{(a) Spatially-averaged phase-matching figure-of-merit at the peak
of the laser pulse, $L_{\mathrm{abs}}/L_{\mathrm{coh}}$ (colour scale
and contour line), for the 11\protect\textsuperscript{th} harmonic,
as a function of peak laser intensity and $\xi_{\mathrm{beam}}$.
Phase matching requires~\citep{Constant1999} $L_{\mathrm{abs}}/L_{\mathrm{coh}}\apprle0.2$.
The markers correspond to the same experimental conditions as displayed
in Fig.~\ref{fig:eta_steady}b. (b) and (c) Measured 11\protect\textsuperscript{th}
and 17\protect\textsuperscript{th} harmonic power versus partial
xenon backing pressure. (d) and (e) The macroscopic response $S_{\mathrm{mac}}$,
which is the measured harmonic power divided by the measured single-atom
response. The overturning peaks in the 9:1 He:Xe gas mix curves in
figures (d) and (e) indicate phase matching.\label{fig:pm_merit}\label{fig:pm_peaks}}
\end{figure*}

Having identified the steady-state plasma as the main limitation for
phase matching, we study its formation dynamics on the relevant time
scale, which is the laser pulse repetition period $\tau_{\mathrm{rep}}=1/f_{\mathrm{rep}}$.
To this purpose, we define two dimensionless parameters: $\xi_{\mathrm{ion}}$
and $\xi_{\mathrm{beam}}$, the number of laser pulses that enter
the gas jet during the time it takes an ion to clear the ion-generation
volume or during the transit time of an atom through the laser beam
volume, respectively. The accumulation of plasma over many pulses
stems directly from the intensity-dependent $\xi_{\mathrm{ion}}$.
However, in order to separate the highly-nonlinear intensity dependence
of plasma accumulation from the effects of other experimental parameters,
it is more convenient to use the intensity-independent parameter $\xi_{\mathrm{beam}}$.

The precise definitions of $\xi_{\mathrm{ion}}$ and $\xi_{\mathrm{beam}}$
are as follows: $\xi_{\mathrm{ion}}=\tau_{\mathrm{ion}}/\tau_{\mathrm{rep}}$,
where $\tau_{\mathrm{ion}}$ is the transit time of an ion through
the half-width at half-maximum (HWHM) of the intensity-dependent ion-generation
volume. This volume is defined by the ionization probability profile
created by a single laser pulse, $\eta_{\mathrm{pulse}}\left(x\right)$,
which is calculated numerically (see Methods). Similarly, $\xi_{\mathrm{beam}}=\tau_{\mathrm{beam}}/\tau_{\mathrm{rep}}$,
where $\tau_{\mathrm{beam}}=\,\sigma_{\mathrm{FWHM}}/v_{\mathrm{gas}}$.
Here, $\sigma_{\mathrm{FWHM}}$ is the full-width at half-maximum
(FWHM) of the intensity profile and $v_{\mathrm{gas}}=\sqrt{5\,R\,T/M_{\mathrm{avg}}}$
is the translational velocity of the gas~\citep{Miller1988} perpendicular
to the laser propagation direction, with $R$ denoting the universal
gas constant, $T$ the gas stagnation (backing) temperature, and $M_{\mathrm{avg}}$
the weighted-average molar mass of the monatomic gas mixture.

The steady-state ionization fraction $\eta_{\mathrm{steady}}$ (spatially-averaged
over the intensity profile) is shown in Figure~\ref{fig:eta_steady}b
as a function of $\xi_{\mathrm{beam}}$ and laser intensity, where
the contour lines correspond to $\xi_{\mathrm{ion}}$. The HHG yield
from a single atom increases with intensity, however, Fig.~\ref{fig:eta_steady}b
shows that higher intensity also corresponds to higher steady-state
ionization fraction. For a fixed intensity, $\eta_{\mathrm{steady}}$
increases with a higher repetition rate or slower gas jet. We note
that decreasing the spot size at fixed peak intensity would also decrease
the steady-state ionization, however, it would reduce the size of
the generation volume, thus preventing a gain in the total harmonic
yield.

The ionization fraction is linked to the phase-matching conditions,
thus determining directly the XUV yield. In HHG, the phase mismatch
is usually expressed as a wave-vector mismatch~\citep{Constant1999,Paul2006,Heyl2012}:
\begin{equation}
\Delta k\approx\rho\left(\left(1-\eta\right)\left|\frac{\partial\Delta k_{n}}{\partial\rho_{n}}\right|-\eta\left|\frac{\partial\Delta k_{p}}{\partial\rho_{p}}\right|\right)-\left|\Delta k_{g}\right|.\label{eq:phase_mismatch}
\end{equation}
Here, $\rho$, $\rho_{n}$, and $\rho_{p}$ are the total, neutral,
and plasma densities, respectively, $\eta$ is the ionization fraction,
$\Delta k_{n}$ and $\Delta k_{p}$ are the wave-vector mismatches
due to neutral and plasma dispersion, respectively, and $\Delta k_{g}$
is the geometric wave-vector mismatch due to the Gouy phase dispersion.
$\Delta k_{g}$ and $\partial\Delta k_{n,p}/\partial\rho_{n,p}$ are
independent of $\rho_{n,p}$. We have not included the intensity-dependent
dipole phase contribution to $\Delta k$, as it is negligible in our
conditions where the generation medium is much shorter than the Rayleigh
length of the generating beam~\citep{Heyl2017}. An optimum harmonic
yield is reached for $\Delta k$ =~0, corresponding to an infinite
coherence length $L_{\mathrm{coh}}=\pi/\Delta k$. From Eq.~(\ref{eq:phase_mismatch})
we see that the neutral and plasma dispersion can compensate the Gouy
phase shift for a certain gas density, but only if the ionization
fraction stays below a critical value. As the HHG radiation is absorbed
by the generation medium, the effective coherence length is limited,
therefore the maximum yield is absorption-limited. A simplified 1D 
analysis~\citep{Constant1999} shows that, for a given absorption length, the harmonic yield
will reach at least half of its optimum when the coherence length
is sufficiently long such that $L_{\mathrm{abs}}/L_{\mathrm{coh}}\le0.2$,
where $L_{\mathrm{abs}}=(\sigma\,\rho)^{-1}$ is the absorption length
with absorption cross section $\sigma$. Since HHG is a highly-nonlinear
process, phase-matching is most important near the peak of the laser
pulse. Figure~\ref{fig:pm_merit}a shows spatially-averaged $L_{\mathrm{abs}}/L_{\mathrm{coh}}$
at the peak of the pulse, simulated for the 11\textsuperscript{th}
harmonic ($\lambda_{11}=97$ nm) under our experimental conditions
(see Methods), on the same axes as those used to display the steady-state
ionization fraction in Fig.~\ref{fig:eta_steady}b. The expected
trend of improved phase matching with lower steady-state ionization,
which is reached for faster gas jet velocities and lower intensities,
is clearly visible.

Plasma accumulation would be avoided in the single pulse regime, $\xi_{\mathrm{ion}}\le1$,
when a significant part of the plasma clears the generation volume
between consecutive pulses. This has not been the case for previous
HHG work at $f_{\mathrm{rep}}\gg10\text{ MHz}$. In order to achieve
single-pulse conditions in our experiment, we increase the gas velocity
by increasing its temperature or by seeding the heavy generator gas
(Xe) in a light carrier gas (He) (i.e., decreasing $M_{\mathrm{avg}}$)~\citep{Miller1988,Mills2012,Carstens2016}.
Under our experimental conditions (77 MHz, focal spot size $\sigma_{\mathrm{FWHM}}=41\;\mathrm{\mu m}$,
$I_{\mathrm{peak}}\approx5\times10^{13}\;\mathrm{W}/\mathrm{c}\mathrm{m}^{2}$)
we expect to transition into the single pulse regime when the helium
fraction is about 80\%, corresponding to a gas speed of 648~m/s (compared
to 305~m/s for pure Xe) at room temperature (see Figs.~\ref{fig:eta_steady}b
and \ref{fig:pm_merit}a). The ionization potential of He (24.6~eV)
is much higher than that of Xe (12.13~eV), thus at our laser intensity,
helium does not contribute to XUV emission and does not add any more
plasma. We note that He dispersion is comparable to Xe dispersion
in a 9:1 He:Xe gas mix, and XUV absorption in He is negligible for
harmonic orders $\leq21$.

\begin{figure*}[t]
\begin{centering}
\includegraphics[width=1\textwidth]{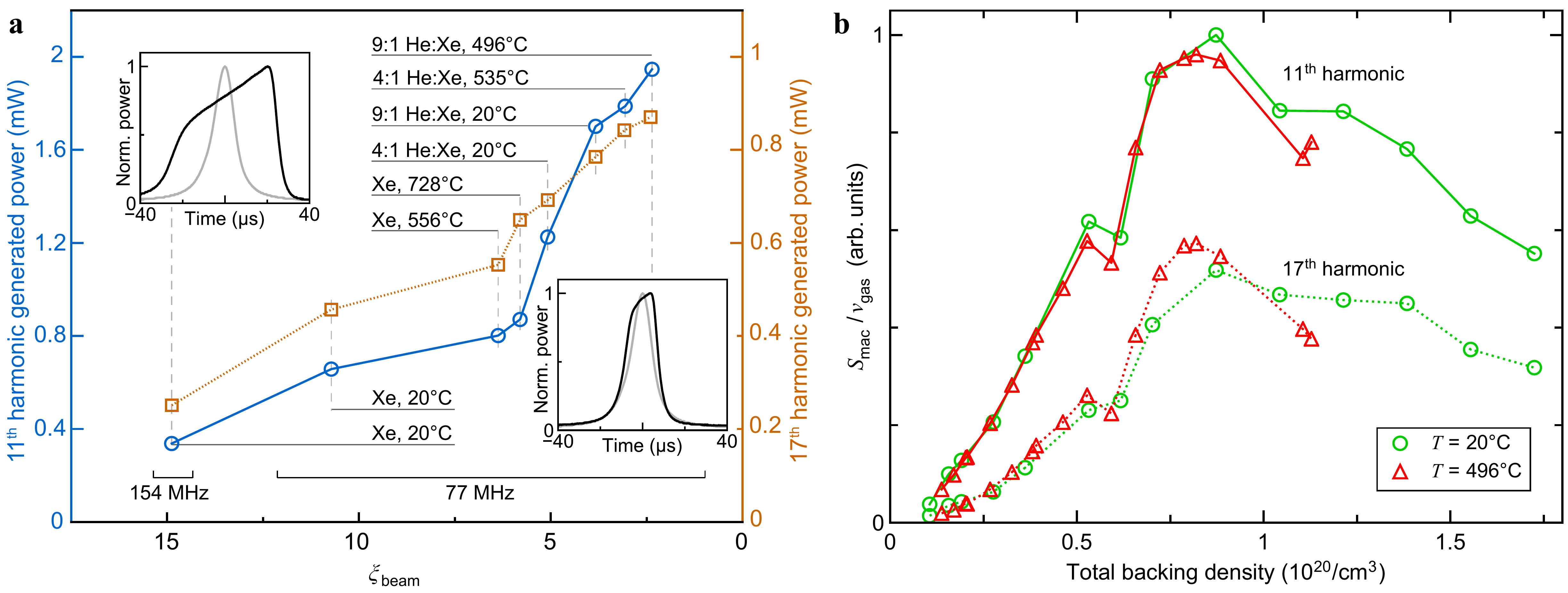}
\par\end{centering}
\caption{(a) Experimentally generated 11\protect\textsuperscript{th} and 17\protect\textsuperscript{th}
harmonic power as a function of $\xi_{\mathrm{beam}}$ for experimental
conditions with different repetition rates, gas mixes, and gas temperatures.
Each point corresponds to the peak of a pressure-curve analogous to
Fig.~\ref{fig:pm_peaks}b and \ref{fig:pm_peaks}c. The insets show
intracavity laser power while sweeping over the cavity resonance with
gas (black) and without gas (grey) for the smallest and largest $\xi_{\mathrm{beam}}$.
(b) Pressure scan of the macroscopic response $S_{\mathrm{mac}}$
scaled by the calculated gas terminal velocity $v_{\mathrm{gas}}$
for the 9:1 He:Xe gas mixture with the nozzle unheated ($20^{\circ}\mathrm{C}$)
and heated (estimated gas temperature ${\sim}$$496^{\circ}\mathrm{C}$).
The horizontal axis is transformed from pressure and temperature to
density assuming ideal gas behaviour.\label{fig:trend}\label{fig:heated}}
\end{figure*}

We perform HHG in an enhancement cavity (see Fig. 1 in Ref. \citen{Cingoz2012}),
where the driving laser power is enhanced by a factor of $\text{\ensuremath{\sim}200}$
at repetition rates of 154~MHz or 77~MHz via pulse picking (see
Methods for details). Reducing the plasma level in the HHG generation
volume comes with additional benefits for intracavity HHG\@. In an
enhancement cavity, the plasma's dispersion limits the power build-up
in the cavity, thus restricting the focus intensity~\citep{Allison2011,Carstens2016,Carlson2011}.
The dependence of $\eta_{\mathrm{steady}}$ on laser intensity causes
optical bistability and coupling to higher order transverse modes
due to plasma lensing, which also limit the intracavity focal intensity
\citep{Yost2011,Mills2012,Allison2011,Carlson2011}.

Figures~\ref{fig:pm_peaks}b and~\ref{fig:pm_peaks}c show the experimentally
generated power $S_{q}$ of the harmonic order $q$ = 11 and 17 ($\lambda_{17}=63$
nm), as a function of partial xenon backing pressure, for the case
of a pure xenon jet and a 9:1 He:Xe gas mixture. Increasing the gas
density increases the dispersive and nonlinear effects of the plasma
in the enhancement cavity~\citep{Allison2011,Carlson2011,Carstens2016},
leading to a decreased intracavity intensity $I$ as the backing pressure
is increased. In order to remove the ambiguity between the effects
of phase matching and intracavity intensity on harmonic yield, we
measured the nonlinear intensity-dependence of the single atom response
$I^{N_{q}}$ for each of the two harmonics (see Methods for details).
In Figs.~\ref{fig:pm_peaks}d and~\ref{fig:pm_peaks}e we plot the
harmonic power divided by the single atom response, $S_{q,\mathrm{mac}}=S_{q}/I^{N_{q}}$,
thus we remove the effect of the variation in generating intensity.
For each of the two harmonics, we observe a clear peak for the case
of the gas mix, as expected for phase-matched and absorption-limited
generation {[}see Eq.~(\ref{eq:phase_mismatch}){]}, while for pure
xenon we obtain a saturation behaviour with no discernible peak. These
results indicate that phase matching is reached when the gas mix is
used. 

There are three other effects that could explain a peak in $S_{\mathrm{mac}}$
as a function of pressure, which are excluded in our experiment. First,
absorption of the generated harmonics behind the generation medium.
This can be excluded as the chamber background pressure is kept below
$2\times10^{-2}\,\textrm{mbar}$, and the Rayleigh length is much
greater than the effective medium length defined by the gas distribution.
Second, a possible reshaping of the laser pulse upon propagation through
the nonlinear medium would depend on the gas density and could cause
unfavourable phase matching at high densities. However, significant
reshaping effects would also affect the harmonics generated with pure
Xe. Third, clustering at high densities could influence the HHG process.
In order to exclude effects due to clustering, we heated our nozzle
to $560^{\circ}\mathrm{C}$, leading to estimated gas temperatures
of ${\sim}$$496^{\circ}\mathrm{C}$ for the 9:1 He:Xe mixture. This
corresponds to a ${\sim}$$1.6\times$ higher gas jet speed due to
heating. We note that we empirically observed a linear scaling of
the HHG yield with gas jet speed when $\xi_{\mathrm{beam}}<6$ (see
Fig.~\ref{fig:trend}a). In this range the steady-state plasma is
mostly, but not completely, avoided, so increasing the gas jet speed
still gradually reduces the steady-state plasma. The effect of this
$\eta_{\mathrm{steady}}$ reduction on harmonic yield can thus be
modelled with a first order approximation, i.e., a linear scaling.
It is important to note that this scaling does not affect any conclusions
regarding the macroscopic generation conditions, since a change in
the macroscopic conditions would change the harmonic yield dependence
on density, not just the overall level. The measured harmonic power
scaled according to the single atom response and the gas jet terminal
velocity, $S_{q,\mathrm{mac}}/v_{\mathrm{gas}}$, is shown in Fig.~\ref{fig:heated}b
as a function of the backing gas density, for both the unheated and
heated nozzle case with 9:1 He:Xe gas mix. The two data sets match
very well, showing that in both cases the dispersion and absorption
properties of the generating medium conditions are the same. This
indicates that cluster formation, which is strongly dependent on temperature~\citep{Tisch1997},
does not play a role here. This conclusion is also supported by a
recent investigation of xenon clustering in HHG, in which such effects
were only observed at a large distance from the nozzle where the clusters
contain at least $10^{4}$ atoms~\citep{Li2016}, while here we perform
HHG within 1\textendash 2 nozzle diameters from the orifice. 

\begin{figure}[tb]
\begin{centering}
\includegraphics[width=0.5\textwidth]{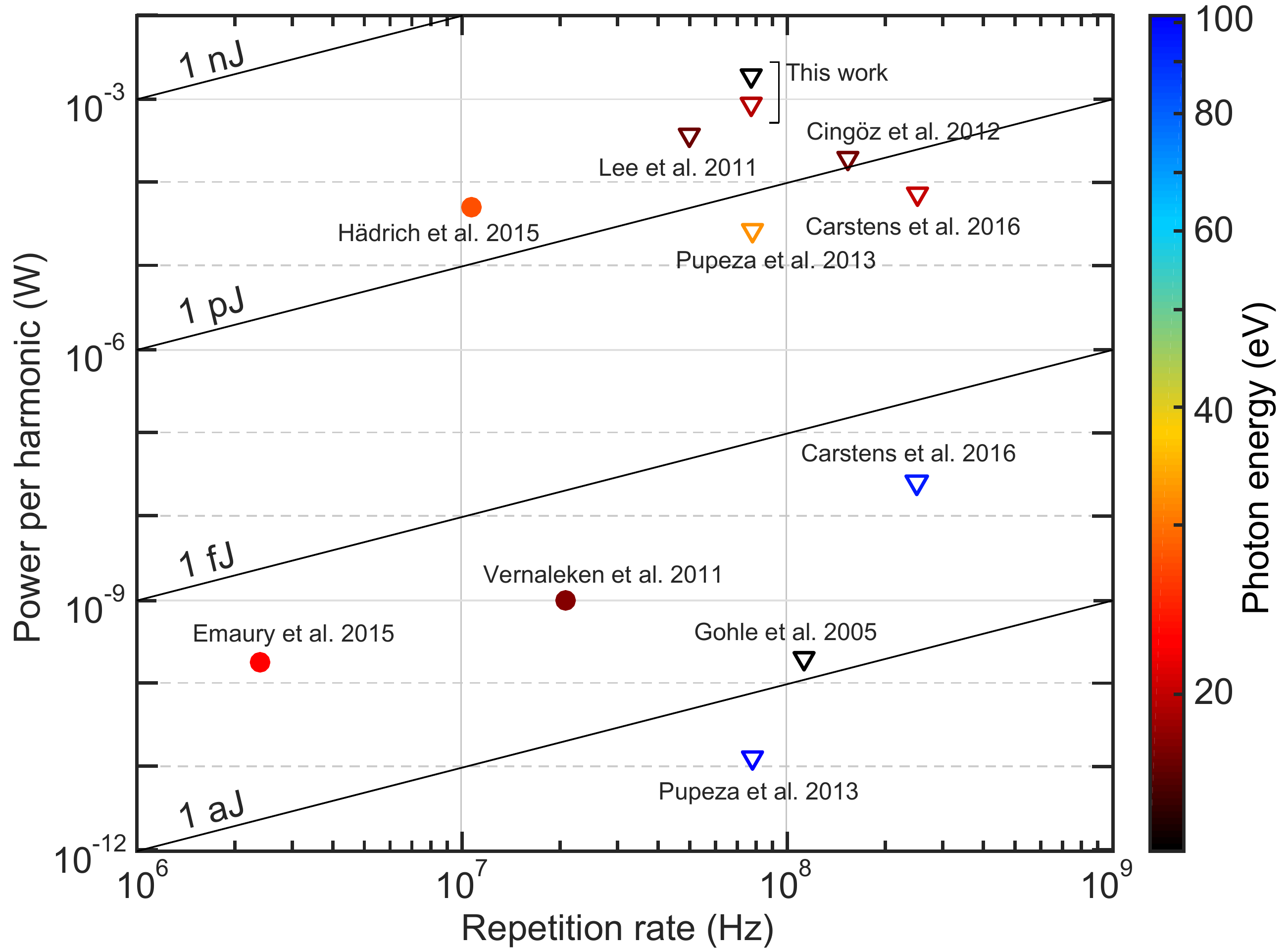}
\par\end{centering}
\caption{Overview of experimentally generated power per harmonic above 10~eV
at repetition rates above 1~MHz, in single-pass configuration (circles)~\citep{VernalekenOL2011,EmauryOpt2015,Hadrich2015}
and intracavity configuration (triangles)~\citep{Gohle2005,Lee2011,Cingoz2012,Pupeza2013,Carstens2016}.
The photon energy is indicated by marker colour. For the cases where
the cavity out-coupling efficiency was not reported, we assume 10\%
and 5\% for Brewster plate out-coupler \citep{Lee2011} and hole mirror
\citep{Carstens2016}, respectively.\label{fig:overview}}
\end{figure}

We experimentally studied the dependence of generated harmonic power
on $\xi_{\mathrm{beam}}$, using different gas mixes and temperatures,
displayed in Fig.~\ref{fig:trend}a. Both the 11\textsuperscript{th}
and 17\textsuperscript{th} harmonic power monotonically increase
as $\xi_{\mathrm{beam}}$ decreases. For both harmonics, the power
increases with a roughly-constant slope as $\xi_{\mathrm{beam}}$
is reduced to $\textrm{\ensuremath{\sim}}6$. This trend stems from
the increased number of neutral xenon atoms available for HHG and
a slight improvement of phase matching. Beyond this point, the slope
increases dramatically. We attribute this dramatic increase to the
onset of significant phase matching due to the transition into the
single-pulse regime. Indeed, as seen in Fig.~\ref{fig:eta_steady}b,
$\xi_{\mathrm{beam}}\approx6$ corresponds to $\xi_{\mathrm{ion}}\approx1$.
As explained above, the continued gradual reduction of the steady-state
plasma is the reason the measured harmonic yield does not saturate.
Another indication that the steady-state plasma is decreasing is the
mitigation of plasma-induced cavity bistability \citep{Allison2011,Carlson2011,Yost2011}
for lower $\xi_{\mathrm{beam}}$, indicated by the narrower and nearly
Lorentzian intracavity power curve measured by sweeping the cavity
length, shown in the insets of Fig.~\ref{fig:trend}a. To connect
the experimental results of Fig.~\ref{fig:trend}a with our calculation
of steady-state ionization and phase matching, we placed markers on
Figs.~\ref{fig:eta_steady}b and~\ref{fig:pm_merit}a, which correspond
to pressure-optimized 11\textsuperscript{th}-harmonic conditions
for various $f_{\mathrm{rep}}$, $v_{\mathrm{gas}}$, and $\sigma_{\mathrm{FWHM}}$.
Fig~\ref{fig:eta_steady}b shows that in the cases of a pure xenon
generation medium and either 154~MHz or 77~MHz repetition rate,
the single pulse regime is not reached, as opposed to a 9:1 He:Xe
gas mix at 77~MHz repetition rate at room temperature and heated.
Correspondingly, Fig.~\ref{fig:pm_merit}a predicts a significant
improvement in phase matching for the 11\textsuperscript{th} harmonic
for the last two cases, as corroborated by the experimental results
in Figs.~\ref{fig:pm_peaks}d and \ref{fig:trend}a. Finally, we
point out the fact that for the highest gas speeds we have achieved
record power levels of $\sim2$ mW and $\sim0.9$ mW for the 11\textsuperscript{th}
and 17\textsuperscript{th} harmonic, respectively (see Fig.~\ref{fig:overview}).
These correspond to intracavity conversion efficiencies of $1.8\times10^{-7}$
and $8.5\times10^{-8}$, which are comparable to those obtained with
single-pass phase-matched HHG using a similar generating wavelength
at $f_{\mathrm{rep}}=10.7\,\mathrm{MHz}$ \citep{Hadrich2015}. 

In conclusion, we have demonstrated phase-matched high repetition
rate ($\gg\!10$~MHz) HHG for the first time, achieving beyond mW
power levels per harmonic order, bringing the generated XUV brightness
to a similar level obtained from synchrotron sources \citep{Allison2011}.
Our results not only set a new power record for HHG-based XUV sources
in general (including low-repetition rate systems), they also open
the door to direct frequency comb spectroscopy in few-electron systems
such as He \citep{Eyler2008} and highly charged ions \citep{LopezJPCS2015}.
We have shown that steady-state plasma mitigation is possible and
critical for phase-matching high repetition rate HHG, and that a simple
model of plasma motion is sufficient to capture all of the important
dynamics involved in plasma accumulation and predict the conditions
required for phase matching. The universal physical insight we provide
will be indispensable for phase matching and power scaling HHG driven
by emerging laser technology with shorter pulses, higher repetition
rate and higher power.

\section*{Methods}

\subsection*{Experimental Apparatus}

The main experimental system is described in detail elsewhere~\citep{Cingoz2012,Benko2014};
here we provide a brief overview. The laser is a Yb:fibre amplified
frequency comb delivering 120~fs pulses, spectrally centred at 1070~nm~\citep{Ruehl2010}.
By optionally inserting an electro-optic pulse picker before the final
fibre amplifier stage, we operate with an amplifier-saturated output
power of 80~W at either $f_{\mathrm{rep}}=154$~MHz or 77~MHz.
The 3.9~m roundtrip ring cavity (single-pulse resonant for 77~MHz;
two-pulse resonant for 154 MHz) is found to give the same HHG performance
at 154~MHz as a 1.54~m roundtrip cavity (single-pulse resonant for
154~MHz) with the same finesse and focal spot size. In order to maintain
the same peak intensity at the same average power for both repetition
rates, the cavity is operated with a focal spot size of $\sigma_{\mathrm{FWHM}}=29\;\mathrm{\mu m}$
at $f_{\mathrm{rep}}=154\;\mathrm{MHz}$, and this is increased by
$\sqrt{2}$ to $\sigma_{\mathrm{FWHM}}=41\;\mathrm{\mu m}$ at $f_{\mathrm{rep}}=77\;\mathrm{MHz}$.
The focal spot size is determined experimentally from the frequency
spacing of cavity-swept high-order modes~\citep{Siegman_book}. The
gas jet is injected into the cavity focus with a home-made~\citep{Heyl_nozzle}
fused-silica nozzle with a 36~mm temperature-controlled section wrapped
in resistive heater wire, followed by a 50~$\text{\ensuremath{\mu}m}$
diameter orifice at the tip. Differential gas pumping is maintained
by a 1.5~mm diameter orifice gas-catch assembly placed at a distance
of ${\sim}$0.5~mm from the nozzle orifice.

The generated XUV harmonics are outcoupled from the cavity by a flat
mirror with a nano-etched surface~\citep{Yost2008} which acts as
a grating for the XUV with an outcoupling efficiency of 6.55\% and
10.52\% for the 11\textsuperscript{th} and 17\textsuperscript{th}
harmonics, respectively, calculated from the groove depth and period
measured with an atomic-force microscope. The two selected harmonics
are each directed to their detectors via one grazing-incidence reflection
on bare-gold mirrors. The 11\textsuperscript{th} harmonic is measured
with a NIST-calibrated $\mathrm{Al_{2}O_{3}}$ windowless photoemissive
detector~\citep{Vest1999}, and the 17\textsuperscript{th} harmonic
is measured using an aluminium-foil-coated silicon photodiode (Opto
Diode AXUV100Al). The Si photodiode is calibrated against the NIST
photoemissive detector by measuring the 17\textsuperscript{th} harmonic
power with both detectors sequentially, under easily-repeatable conditions
(unheated pure Xe). We estimate the upper bound of the uncertainty
in the generated harmonic power measurement to be $\text{\ensuremath{\pm}7\%}$.
All measurements reported in this article were taken while sweeping
the cavity length across its resonance at a rate much slower than
the cavity lifetime.

\subsection*{Ionization and Phase-matching Theory}

The theoretical ionization fraction is calculated using the modified
PPT (Perelomov-Popov-Terent'ev) ionization model~\citep{Zhao2016}.
The ionization probability during a single laser pulse is $\eta_{\mathrm{pulse}}(x,t)=1-\exp\!\!\left[-\int_{-\tau_{\mathrm{rep}}/2}^{t}\;w(x,t^{\prime})\,\mathrm{d}t^{\prime}\right]$,
where $w(x,t)$ is the PPT ionization rate. Correspondingly, the single-pulse
ionization probability profile is $\eta_{\mathrm{pulse}}(x)=\eta_{\mathrm{pulse}}(x,\tau_{\mathrm{rep}}/2)$.

The total ionization fraction at position $x$ along the gas propagation
direction, and time $t$ during the laser pulse, is calculated as:

\begin{equation}
\eta(x,t)=\eta_{\mathrm{steady}}(x)+\left[1-\eta_{\mathrm{steady}}(x)\right]\,\eta_{\mathrm{pulse}}(x,t)\,,\label{eq:ionization_fraction}
\end{equation}
where $\eta_{\mathrm{steady}}(x)$ is the steady-state ionization
fraction built up by all previous pulses, and $\left[1-\eta_{\mathrm{steady}}(x)\right]$
is the steady-state neutral fraction. The steady-state ionization
fraction is found by starting with $\eta_{\mathrm{steady}}(x)=0$
and recursively solving Eq.~(\ref{eq:ionization_fraction}) after
shifting the ionization profile spatially by $v_{\mathrm{gas}}\,\tau_{\mathrm{rep}}$
for each repetition period. The calculation neglects plasma recombination
and diffusion, i.e., it assumes that the dominant mechanism determining
pulse-to-pulse plasma survival in the focal volume is the forward
motion of the plasma in the gas jet. Inaccuracies in the PPT model
and the effect of spatial averaging are taken into account, as explained
in the next section. As a result, the spatially-averaged steady-state
ionization fraction is presented in colour scale in Fig.~\ref{fig:eta_steady}b.

The neutral-atom absorption and dispersion optical constants used
in the calculation of $L_{\mathrm{abs}}/L_{\mathrm{coh}}$ in Fig.~\ref{fig:pm_merit}a
are taken from refs.~\citep{LHuillier1990,Masunaga1965,Reinsch1985}.
The coherence length is calculated using Eq.~(\ref{eq:phase_mismatch}),
and $\eta$ is taken as the spatially-averaged ionization fraction
at the peak of the laser pulse, calculated using Eq.~(\ref{eq:ionization_fraction}).
The calculation is performed with $\sigma_{\mathrm{FWHM}}=41\;\mathrm{\mu m}$
focal diameter and 2.4~bar partial xenon backing pressure, which
was the optimal experimental pressure for the 11\textsuperscript{th}
harmonic with 9:1 He:Xe mixture and close to optimum for the other
mixtures. The gas density in the focal volume is taken to be 10\%
of the backing density, which is a reasonable estimation for generation
within 1\textendash 2 nozzle diameters of the nozzle exit~\citep{Miller1988,Carlson2011}.
As the neutral phase mismatch depends on the gas mixing ratio, and
the Gouy phase mismatch depends on focal spot diameter, the black
diamond and blue square markers in Fig.~\ref{fig:pm_merit}, representing
data taken at 154~MHz and 77 MHz with pure xenon, respectively, do
not correspond to the same phase-matching conditions as the other
markers. Therefore, the calculated $L_{\mathrm{abs}}/L_{\mathrm{coh}}$
differ by 6\% and 11\%, respectively, from the displayed colour scale
value. However, for completeness and as the error is small, we choose
to include the markers.

\subsection*{Single-Atom Intensity Dependence and Ionization Model}

We measure the intensity-dependence of the single-atom response by
continuously monitoring the intracavity power by leakage through a
cavity mirror, and scanning the cavity input power while measuring
the harmonic power for an extremely-low gas pressure. In particular,
we use an unheated 4:1 He:Xe gas mix and a 9:1 He:Xe gas mix with
${\sim}$860~mbar total backing pressure. In these low-density conditions,
absorption is negligible and $\Delta k$ is dominated by the Gouy
phase mismatch {[}see Eq.~(\ref{eq:phase_mismatch}){]}, so that
phase matching is independent of intracavity power (or intensity).
Furthermore, the low gas density makes the impact of dispersion and
plasma dynamics on the cavity performance negligible. To determine
the intensity-dependence of the single-atom response~\citep{Huillier1992}, $I^{N_{q}}$, with $N_{q}<q$ for high-harmonic order $q$,
we fit the low intensity part of the scan, where $\eta\ll1$, using
$N_{q}$ as a fitting parameter. We find $N_{11}=4.43(7)$ and $N_{17}=4.75(11)$,
respectively, in agreement with past results~\citep{Huillier1992}. 

The PPT ionization model is known to provide the correct trend but
not the correct magnitude of the ionization rate \citep{Zhao2016}.
Additionally, we want to account for the effect of spatial averaging.
Therefore, we use the high-intensity part of the scan, which deviates
from the $I^{N_{q}}$ power law due to significant neutral depletion,
in order to fit a multiplicative factor to the PPT ionization rate.
The harmonic signal is fitted to the temporally-integrated product
$I^{N_{q}}\left(1-\eta\right)^{2}$ (using the values of $N_{q}$
obtained from the low intensity scan), resulting in a global multiplicative
fitting factor of 0.52(4) for $w\left(x,t\right)$. This fitting factor
effectively accounts for PPT inaccuracies, spatial averaging, and
the systematic error in the measured intensity (estimated to be at
most 20\%).

\bibliographystyle{custom_bibstyle}
\bibliography{ICHHG_PM}

\subsection*{Acknowledgements}

This work was supported by the Air Force Office of Scientific Research,
National Institute of Standards and Technology and the National Science
Foundation Physics Frontier Center at JILA (PHY-1734006). C.M.H. was
supported by the Swedish Research Council. K.L.C. acknowledges the
support of the JILA Visiting Fellows Program.

\subsection*{Author contributions}

All authors contributed to the design, planning and execution of the
experiment. G.P., C.M.H., S.B.S., C.B., and J.Y. analysed the data.
All authors contributed to the writing of the manuscript. G.P., C.M.H.,
and S.B.S. contributed equally to this work.

\subsection*{Competing financial interests}

The authors declare no competing financial interests.
\end{document}